\documentclass[prl,aps,reprint,superscriptaddress]{revtex4-1}
\usepackage{marvosym}
\usepackage{graphicx,subfigure}         
\usepackage{amsmath} 
\usepackage{amssymb,stmaryrd}   
\usepackage{color}
\usepackage{wrapfig}
\usepackage{mathtools}
\usepackage{tikz}  
\usetikzlibrary{arrows,shapes,trees,positioning}  
\usepackage{comment} 
\usepackage{units}
\usepackage{siunitx}

\usepackage{hyperref}
\hypersetup{
    colorlinks=true,
    linkcolor=black,
    citecolor=blue,
    urlcolor=blue,
}
\usepackage{ragged2e}

\begin{document}


\title{\textbf{Damping-like Torque in Monolayer 1T-TaS$_2$}}

\author{Sajid Husain}
\affiliation{Department of Material Sciences and Engineering, Uppsala University, Box 534, SE-751 21 Uppsala, Sweden}

\author{Xin Chen}
\affiliation{Department of Physics and Astronomy, Materials Theory, Uppsala University, Box 516, SE-751 20 Uppsala, Sweden}

\author{Rahul Gupta}
\affiliation{Department of Material Sciences and Engineering, Uppsala University, Box 534, SE-751 21 Uppsala, Sweden}

\author{Nilamani Behera}
\affiliation{Department of Material Sciences and Engineering, Uppsala University, Box 534, SE-751 21 Uppsala, Sweden}

\author{Prabhat Kumar}
\affiliation{Department of Thin Films and Nanostructures, Institute of Physics of the Czech Academy of Sciences, Cukrovarnická 10/112, 162 00 Prague, Czech Republic}

\author{Tomas Edvinsson}
\affiliation{Department of Material Sciences and Engineering, Uppsala University, Box 534, SE-751 21 Uppsala, Sweden}

\author{Felipe Garcia Sanchez}
\affiliation{Istituto Nazionale di Ricerca Metrologica (INRIM), Strada delle Cacce 91, 10135 Torino, Italy}

\author{Rimantas Brucas}
\affiliation{Department of Material Sciences and Engineering, Uppsala University, Box 534, SE-751 21 Uppsala, Sweden}

\author{Sujeet Chaudhary}
\affiliation{Thin Film Laboratory, Department of Physics, Indian Institute of Technology Delhi, New Delhi 110016, India}

\author{Biplab Sanyal}
\affiliation{Department of Physics and Astronomy, Materials Theory, Uppsala University, Box 516, SE-751 20 Uppsala, Sweden}

\author{Peter Svedlindh}
\affiliation{Department of Material Sciences and Engineering, Uppsala University, Box 534, SE-751 21 Uppsala, Sweden}

\author{Ankit Kumar}
\affiliation{Department of Material Sciences and Engineering, Uppsala University, Box 534, SE-751 21 Uppsala, Sweden}

\date{\today}

\begin{abstract}

A damping-like spin orbit torque (SOT) is a prerequisite for ultralow power spin logic devices. Here, we report on the damping-like SOT in just one monolayer of the conducting transition metal dichalcogenide (TMD) TaS$_2$ interfaced with a NiFe (Py) ferromagnetic layer. The charge-spin conversion efficiency is found to be 0.25$\pm$0.03 and the spin Hall conductivity (\,2.63 $\times$ 10$^5$ $\frac{\hbar}{2e}$ $\Omega^{-1}$ m$^{-1}$)\, is found to be superior to values reported for other TMDs. The origin of this large damping-like SOT can be found in the interfacial properties of the TaS$_2$/Py heterostructure, and the experimental findings are complemented by the results from density functional theory calculations. The dominance of damping-like torque demonstrated in our study provides a promising path for designing next generation conducting TMD based low-powered quantum memory devices.

\end{abstract}


\maketitle

\section{\label{sec:one}Introduction}
Spin-orbit torques ( \,SOTs)\, induced by spin currents are prerequisite to control the magnetization (${m}$) in next generation non-volatile three terminal memory devices [\onlinecite{emori2013current,safeer2016spin,garello2013symmetry}] spin torque nano-oscillators for microwave assisted switching and neuromorphic computing \cite{chen2016spin}. SOT based magnetic memories are considered to be more reliable by utilizing low energy induced switching of the magnetization in contrast to the low endurance and low speed of two terminal spin transfer torque (STT) based random access memories. A spin current with spin polarization vector $\sigma$ generated by the spin Hall effect (SHE) and/or the Rashba-Edelstein effect (REE) in presence of high spin-orbit coupling (SOC) in a material may give rise to two types of SOTs; damping-like (${\tau_{DL}}={m} \times ({m}\times {\sigma})$) and field-like (${\tau_{FL}}={m}\times {\sigma}$)
torques, and have been reported for a number of heavy metals (HMs) [\onlinecite{garello2013symmetry,kim2013layer,zhang2015role,kurebayashi2014antidamping,manchon2019current}]. In contrast to STT devices where the spin polarization of the charge current passing through the free layer enforce the switching of the magnetization, the physical origin of SOTs is the transfer of spin  and orbital angular momenta through exchange interaction process[\onlinecite{manchon2019current}] the latter via contributions mainly from different d orbitals such as d$_{xy}$, d$_{xz}$, d$_{yz}$, d$_{z^2}$, d$_{x^2-y^2}$ [\onlinecite{mahfouzi2020microscopic}]. The SOT in HMs is reported to be a bulk-like phenomenon, which requires the thickness of the HM layer to be larger than its spin diffusion length in order to produce appreciable torque [\onlinecite{berger2018determination}]. However, it is difficult to control the crystallinity of the HM in the low thickness regime. Recently, very large SOT has been reported in topological insulators (TIs) [\onlinecite{mellnik2014spin},\onlinecite{khang2018conductive}] but the topological surface states are quenched if the TI is deposited next to a metallic ferromagnet and hence a ferromagnetic insulator is required to render high SOTs [\onlinecite{li2019magnetization}], which implies industrial compatibility issues [\onlinecite{pai2018switching}].

To overcome the bulk-like effect in HMs, with perseved industrial compatibility, two-dimensional transition metal dichalcogenides (2D-TMDs) were proposed a few years ago for spintronic applications [\onlinecite{feng2017prospects,husain2018spin,guimaraes2018spin,lv2018electric}]. By replacing the HMs with TMDs, one can anticipate two  positive outcomes for spin devices. Firstly, a pure spin current can be produced by just a monolayer thick TMD without any bulk-like effect. Secondly, being a layered material, it is possible to realize smooth surfaces with atomic-scale flatness, \textit{i.e.}, in the Ångström scale. Although, there are few reports on the observation of SOTs in TMDs [\onlinecite{lv2018electric},\onlinecite{shao2016strong}] they are, however, encountered with the problem of a dominating field-like torque due to their semiconducting nature [\onlinecite{khang2018conductive},\onlinecite{husain2018spin}]. The surface quality of TMD exfoliated films grown by chemical vapour deposition is also compromised due to high roughness and strain [\onlinecite{you2018synthesis},\onlinecite{yun2015synthesis}]. TMD films produced in this way exhibit inhomogeneity and are thus not suitable for spintronic device applications. Thus, we are forced to face two challenges to realise the requirement of dominating damping-like torques in TMDs. One is to grow large area TMDs directly on SiO$_2$ substrates and concomitantly to provide large damping-like torques by using conducting TMD/ferromagnet bilayers. Keeping in mind the growth problem of conducting TMDs, the 1T-tantalum-disulphide (TaS$_2$) system is yet to be explored which can be easily fabricated by sputtering technique along with the distinctive plasma sulphurization process. Being conducting with high SOC [\onlinecite{sanders2016crystalline}], the 1T-TaS$_2$ system also possesses exotic temperature dependent properties owing to several charge density wave (CDW) transitions [\onlinecite{wang2018chemical}], thus contributing with the rich physics of CDWs to the field of SOTs. Previously, researchers have reported the growth of TaS$_2$ by various methods [\onlinecite{wang2018chemical,navarro2016enhanced,fu2016controlled}], which were limited to growth of TaS$_2$ flakes,  which are yet to be explored for spin orbitonic applications.

Here, we report a dominating damping-like torque in conducting TaS$_2$/Permalloy ( \,Ni$_{81}$Fe$_{19}$)\, 
bilayer heterostructures. 
The high quality large area TaS$_2$ monolayer has been prepared by ion-beam sputtering combined with plasma-assisted sulfurization (\,see methods and materials for details in supplementary information S1)\,. 
The damping-like torque has been studied in lithographically patterned bilayer devices of size 100 $\times$20 $\mu$m$^2$ by spin torque ferromagnetic resonance (\,ST-FMR)\,, current induced switching in an in-plane magnetic field and angle dependent planar Hall effect measurements. The damping-like torque efficiency has been evaluated from the change of the effective damping induced by the applied dc-current in ST-FMR measurements. All measurements were performed at room temperature. Micromagnetic simulations corroborate the experimental results for the dc-current dependent effective damping. First principles calculations based on density functional theory envisage the possible source of the damping-like torque in the TaS$_2$/Py heterostructures.

\begin{figure}[h!tbp]
\includegraphics[width=8cm]{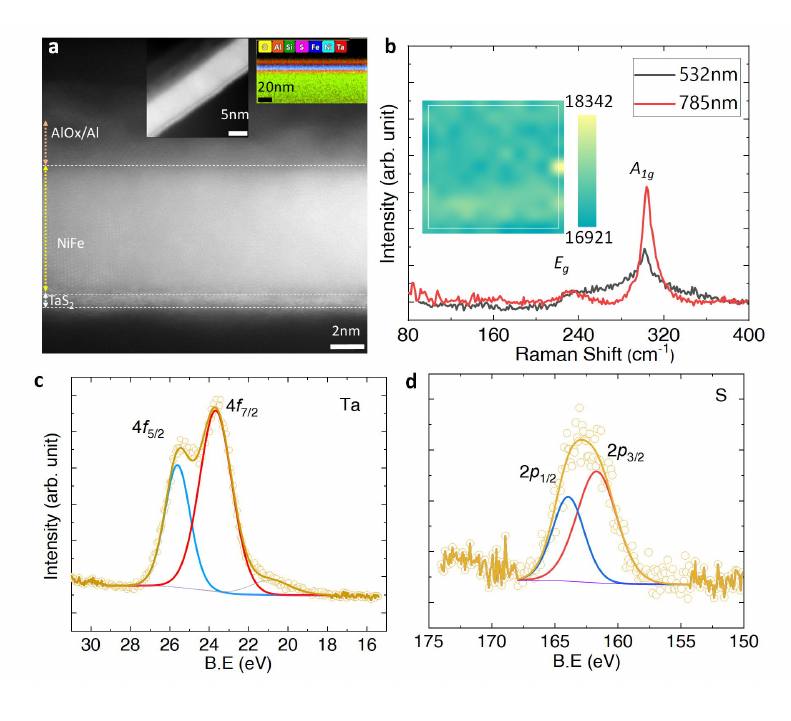}
\caption{\textbf{a,} High resolution cross-sectional TEM image of Py/TaS$_2$/Al heterostructure. Insets: left; low magnification cross sectional image and right; elemental EDX mapping of the trilayer sample. \textbf{b,} Raman spectra recorded using two lasers (532nm and 785nm) and mapping (10 $\times$10$\ \mu m^2$) in inset on single layer TaS$_2$ using a 785nm laser. \textbf{c,} XPS spectra of Ta and S recorded on single layer TaS$_2$ sample.}\label{fig:fig1}
\end{figure}

\begin{figure*}[h!tbp]
    \centering
    \includegraphics[width=15cm]{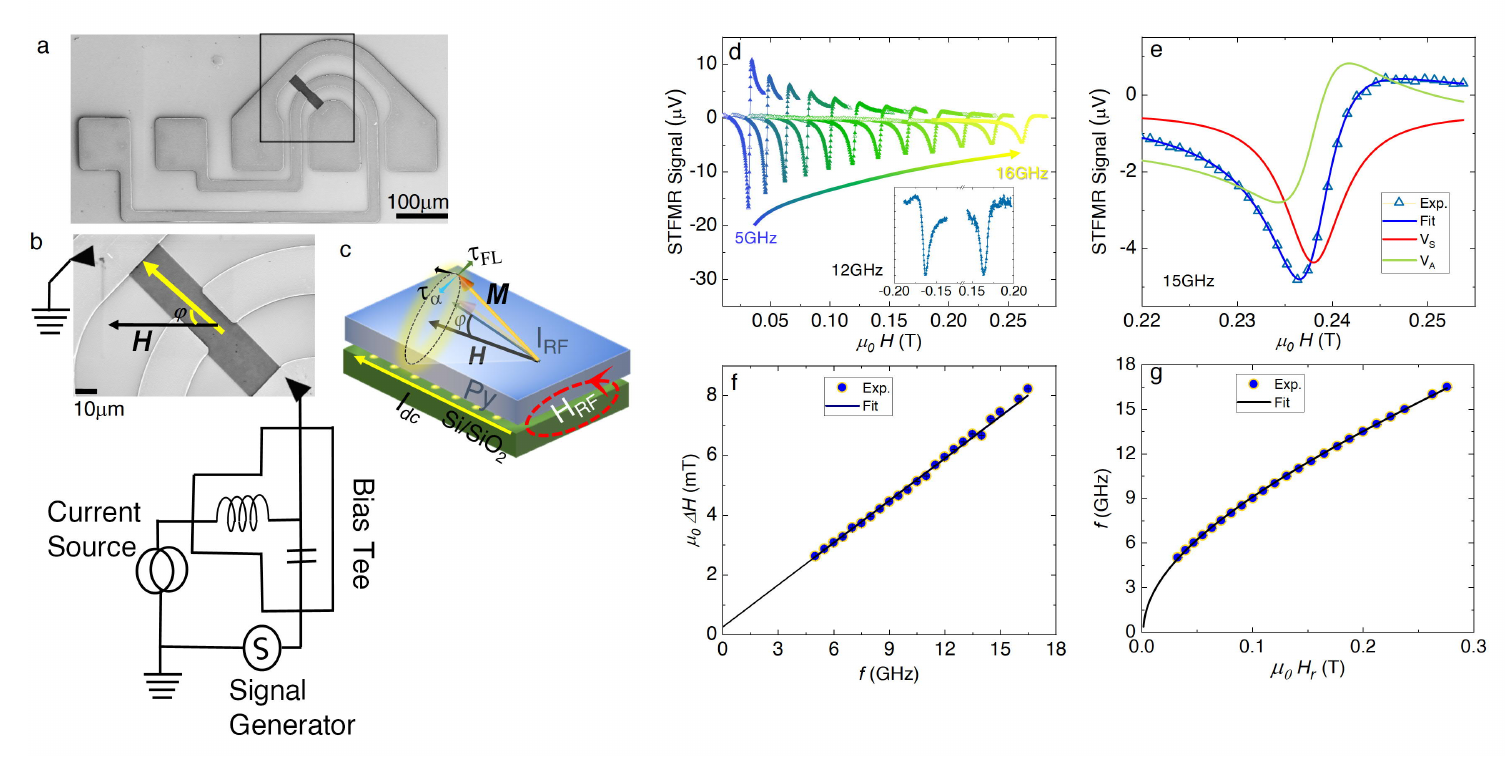}
    \caption{\textbf{a,} Scanning electron microscopic (SEM) image of Py/TaS$_2$ device. \textbf{b,} High magnification SEM image (area indicated by square in a) showing the ST-FMR measurement circuit. \textbf{c,} Layer schematic shown with torques acting on the magnetization. \textbf{d,} ST-FMR spectra recorded at various frequencies in the range of 5-16GHz. Inset; ST-FMR spectra at 12 GHz recorded in opposite magnetic field directions. \textbf{e,} ST-FMR spectrum (recorded at 15GHz) fitted separately with symmetric and anti-symmetric Lorentzian functions. \textbf{f}, and \textbf{g}, linewidth and resonance field versus frequency plots, respectively.}
\end{figure*}
\section{\label{sec:two}
Monolayer characteristics}
Figure 1 shows the transmission electron microscopy ( \,TEM) \, cross sectional image of the Py/TaS$_2$/Al heterostructure.
The thickness of the individual layers are found to be similar to the nominal ones. Notably, the thickness of the TaS$_2$ layer is found to be equivalent to one monolayer. 
In the left inset, the large scale TEM image shows a uniform film and sharp interface of TaS$_2$ in contact with Py layer. Elemental mapping of the stack also supports the uniform growth of all layers in the stack used for device preparation and confirms its composition without interface diffusion or any other impurities. 
It also confirms that our ferromagnet layer shows less affinity to sulfur because metals having large affinity to sulfur can degrade the 2D characteristics of TMDs [\onlinecite{wu2019visualizing}]. Figure 1(b) shows the room temperature Raman spectra recorded on a single layer TaS$_2$ film using two different lasers. Strong fundamental peaks are observed at 305 cm$^{-1}$ and 231 cm$^{-1}$ corresponding to the 1T-TaS$_2$ phase [\onlinecite{hirata2001temperature}]. The uniformity of the film can be seen in the Raman mapping as recorded around the most intense Raman peak (shown in the inset). Elemental analysis has been performed by X-ray photoelectron spectroscopy (XPS) as presented in Figs. 1(c) and 1(d) for Ta and S, respectively. Observed peaks are de-convoluted into the two spin-orbit split peaks which confirm the TaS$_2$ formation without residual phases [\onlinecite{tison2004x},\onlinecite{zeng2014growth}]. Further, surface topography and step height scans were also recorded using atomic force microscopy and confirm the monolayer thickness and the smooth interface with Ångström scale flatness of the TaS$_2$ layer (see Supplementary information S4).
\section{\label{sec:three}Spin torque ferromagnetic measurements }
The magnitude of the SOT efficiency governed by the spin torque efficiency ( \, $\theta_s$) \, was measured using ST-FMR on the photo-lithographically patterned Py/TaS$_2$ device of size 100$\times$20$\mu$m$^2$. The applied field makes an angle of 45$^\circ$ with respect to the current as shown in Fig. 2(a) (scanning electron microscopic (SEM) image of the device). Note that the top Al layer has been removed during the device fabrication process to avoid current shortening from Al (see growth section for details). A SEM image of the measurement circuit is shown in Fig. 2(b). A schematic of the torques acting on the magnetization due to the microwave current I$_{RF}$ is shown in Fig. 2(c). The ST-FMR measurements were performed in field-sweep mode in the frequency range of 5-16 GHz. We have used a lock-in detection technique with an IRF current frequency modulation of 1000 Hz at 9 dB microwave power (see Supplementary information S4 and S5 for details of the measurements and calibration).

The rf current generates an Oersted field as well as spin-orbit torques in presence of the magnetic field and acts as torques on  the magnetization of Py. The I$_{RF}$ induced torque acting on the Py layer generates a sustained precession of the magnetization, which mixes with the anisotropic magnetoresistance and spin Hall magnetoresistance of Py creating a dc mixing voltage V$_{mix}$. This rectified mixing voltage provides the information of the material parameters and torques acting on the magnetization, which is written as [\onlinecite{kumar2017spin}], $V_{mix}=V_0(Sfs+Afa)$. Here, V$_0$ is the amplitude of the mixing voltage and $fs$ and $fa$ are symmetric and antisymmetric Lorentzian functions, respectively. $S=\hbar J_s^{rf}/2e\mu_0 M_st_{Py}$ and $A=H_{RF}\sqrt{(1+(\mu_0M_{eff}/H_r)}$ are symmetric and antisymmetric weight factors, respectively, where $\mu_0$, $e$, $t$, $J_S^{rf}$, $M_S$, $H_{RF}$, $H_r$ and $M_{eff}$ are the magnetic permeability in free space, electronic charge, thickness of ferromagnet layer, rf-spin current density, saturation magnetization, microwave field, resonance field and effective magnetization, respectively.

Figure 2 (d) shows the ST-FMR spectra together with fits using the equation for V$_{mix}$,  which give the line-shape parameters. The ST-FMR spectra for positive and negative magnetic field scans are shown in the inset (at 12 GHz). It is to be noted that the peak changes its sign on changing the direction of the external magnetic field indicating a damping-like torque ($\tau_{DL}$) and ruling out the possibility of a dominating Oersted field generated torque ($\tau_{FL}$). The symmetric and anti-symmetric amplitudes have been separately fitted to the spectra; an example for the spectrum recorded at 15 GHz is shown in Fig. 2(e). The effective damping ($\alpha_{eff}$) of the TaS$_2$/Py bilayer is evaluated by fitting the $\mu_0$ $\Delta$ H versus $f$ data (as shown in Fig. 2(f)) using the equation $\mu_0 \Delta H=\mu_0 \Delta H_0+2\alpha_{eff} \omega / \gamma $, where $\mu_0 \Delta H_0$ is the linewidth contribution from inhomogeneity in the magnetic film and $\omega (=2\pi f)$ is the angular microwave frequency. From the fitting, the values of effective damping is found to be 0.0067$\pm$0.0007, which matches well with the bulk value of Py [\onlinecite{zhao2016experimental}]. The inhomogeneous linewidth is found be 0.20$\pm$0.02mT, which is quite small and indicative of a smooth and clean interface of the Py/TaS$_2$ heterostructure. Further, $\mu_0 M_{eff}$ and the anisotropy field ($\mu_0 H_K$) values have been calculated by fitting the $f$ versus $\mu_0 H_r$ data to the Kittel equation, $f$=$\frac{\mu_0 \gamma}{2\pi}$ $\sqrt{(H_r+H_K)(H_r+H_K+M_{eff})}$ , yielding 0.902$\pm$0.004T and 1.8$\pm$0.4mT, respectively. The values of $\mu_0 M_s$ of TaS$_2$/Py is measured using a QD-MPMS setup and found to be 1.00$\pm$0.02 T (see supplementary information S5 for details), which is consistent with the $\mu_0 M_{eff}$ value extracted from the ST-FMR results considering the out-of-plane anisotropy field contribution to the effective magnetization.

\begin{figure*}[h!tbp]
\centering
\includegraphics[width=17cm]{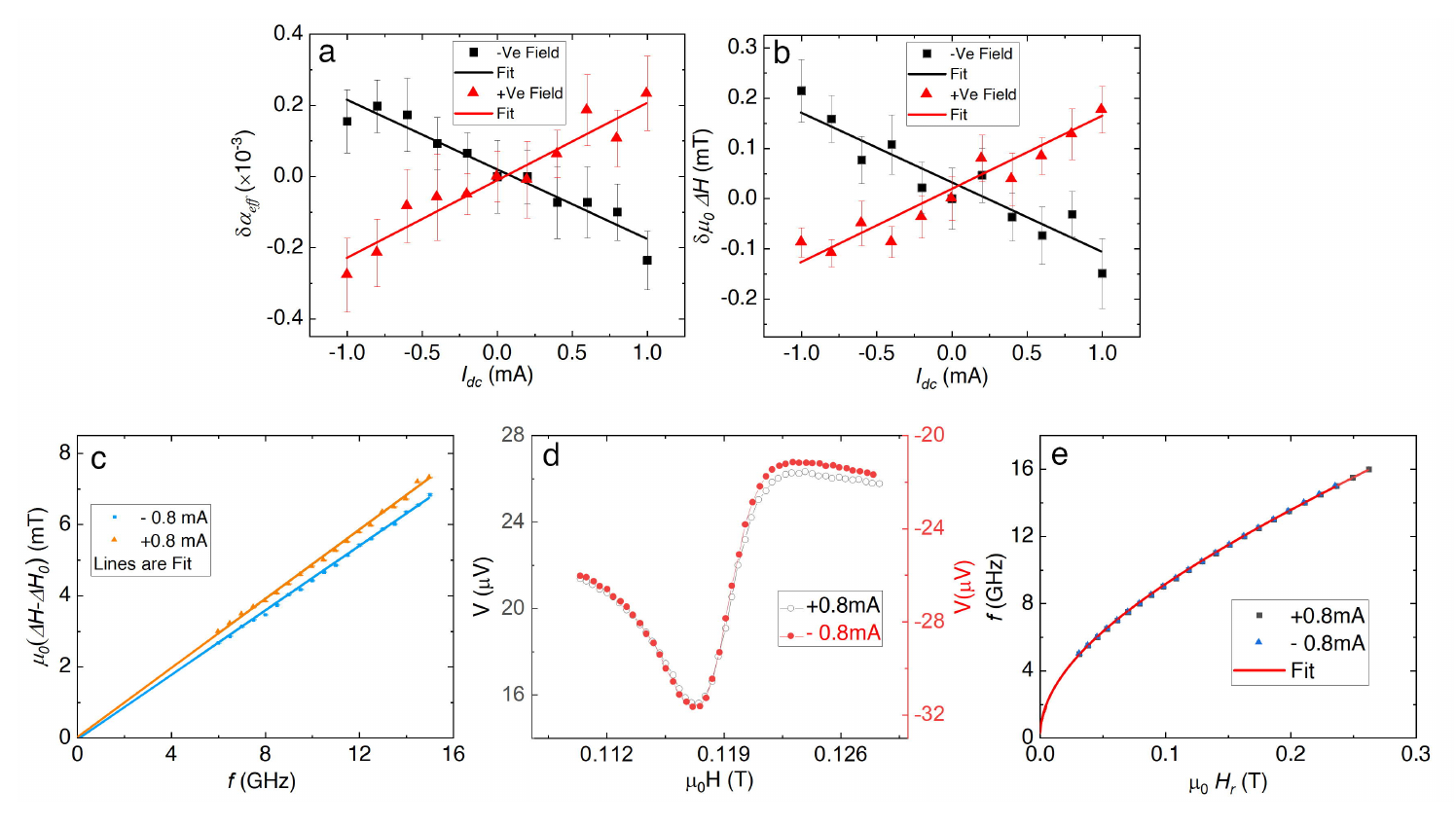}\\
\caption{\textbf{a,} dc-current induced changes of effective damping in Py/TaS$_2$ bilayer for positive and negative field directions. \textbf{b,} dc-current induced linewidth changes for positive and negative magnetic field. \textbf{c,} Linewidth versus frequency. \textbf{d,} ST-FMR spectra recorded at 10GHz and e, frequency versus resonance field measured at +0.8mA and -0.8mA dc-current applied to the device in positive magnetic field. }\label{fig:fig3}
\end{figure*}

From the line-shape parameters, the value of the spin torque efficiency $\theta_s$ is evaluated using the standard line-shape analysis method [\onlinecite{liu2011spin}], and found to be 0.023$\pm$0.01. However, in this method it is assumed that the symmetric component is purely from a damping-like torque, disregarding a possible contribution from spin pumping due to the inverse spin Hall effect and can therefore yield erroneous values for the SOT efficiency [\onlinecite{tserkovnyak2002enhanced,demasius2016enhanced,saitoh2006conversion}]. Concomitantly, the antisymmetric component is considered as an Oersted field generated torque component but it is again a naive approximation [\onlinecite{demasius2016enhanced}]. Moreover, line-shape analysis also shows a frequency dependency [\onlinecite{liu2011spin}], which may lead to the wrong estimation of the effective spin torque efficiency. Hence, to determine a reliable value of the effective spin torque efficiency and evidence of damping-like torque, we use the so called damping modulation scheme by applying a dc-current during the ST-FMR measurement, where spin pumping due to the inverse spin Hall effect and field-like contributions are insignificant.

\section{\label{sec:four}Current induced modulation/changes of effective damping in T\lowercase{a}S$_2$/P\lowercase{y} device}
In this method, the dc-current induced non-equilibrium spin accumulation at the interface, resulting due to the spin Hall effect in TaS$_2$ , acts as a torque on the Py magnetization resulting in a change of the effective damping as described by[\onlinecite{liu2011spin},\onlinecite{demasius2016enhanced}],
$\delta \alpha_{eff}=\alpha_{eff}(I_{dc} \neq 0)-\alpha_{eff}(I_{dc} = 0)= \frac{\sin(\varphi)} {(H_r+0.5M_{eff})\mu_0 M_s t_{py}} \frac{\hbar}{2e} J_{S,dc}$.
The effective spin torque efficiency using the dc-current induced change of the damping is defined as the ratio of spin to charge current density, $\theta_S^\alpha = J_{S,dc}/J_{C,dc}$. Here, $J_{C,dc}=\frac{I_{dc}}{A_d} \frac{R_{Py}}{(R_{Py}+R_{TaS_2}}$, where R$_{Py}$, R$_{TaS_2}$ and A$_d$ are the resistances of permalloy and TaS$_2$, respectively, and the cross sectional area of the device. $\varphi$ is the angle between the magnetization and the applied field which is 45$^\circ$ in our case. Figure 3(a) shows the change/modulation of effective damping as a function of the dc-current (I$_{dc}$). For comparison, the $\delta \alpha_{eff}$ values are plotted for the two directions of magnetic field scan. The corresponding changes of $\mu_0\Delta H$ with dc-current, \textit{i.e.,} $\delta \mu_0\Delta H$ are shown in Fig. 3(b). The slopes of the $\delta \alpha_{eff}$ versus I$_{dc}$ curves for the two field directions are almost equal, which confirm that the damping-like torque acting on the magnetization in our Py/TaS$_2$ bilayer is due to the SHE generated spin current. The slope of the change in $\alpha_{eff}$ with dc-current ($\delta \alpha_{eff}/\delta I_{dc}$) is 2.17$\pm$0.21$\times$10$^{-4}$/mA and 1.95$\pm$0.17$\times$10$^{-4}$/mA for positive and negative applied fields, respectively. Using the measured resistance values of Py (225 $\Omega$) and TaS$_2$ (952 $\Omega$) (see Supplementary information S6 for details of the resistance measurements) in the equation for J$_{C,dc}$, the $\theta_S^\alpha$ value is found to be 0.25$\pm$0.03. Within the experimental uncertainty, the values of the spin torque efficiency are same for both positive and negative field scans. The obtained value is better than values reported for other TMDs [\onlinecite{zhao2020observation,shi2019all,xu2020high}]. The intercept with the current axis is known as the critical current density for auto-oscillations and is estimated to be 5.13$\times$10$^{10}$ A/m$^2$. The spin Hall conductivity ($\sigma_S=\sigma_{dc}\times \theta_S^\alpha$) in units of $\hbar$/2e is found to be 2.63$\times$10$^5$ $\hbar$/2e $(\Omega m)^{-1}$, which is about 50 times smaller than the value reported for the topological insulator Bi$_{0.9}$Sb$_{0.1}$/MnGa [\onlinecite{khang2018conductive}], 100 times larger than for the field-like torque dominated semiconducting TMDs MoSe$_2$ and WS$_2$ [\onlinecite{shao2016strong}], 10 times larger than for the conducting TMD NbSe$_2$[\onlinecite{guimaraes2018spin}]  and comparable to that of the Pd$_{1-x}$Pt$_x$ alloy reported to exhibit a strong damping-like SOT [\onlinecite{zhu2019strong}]. Topological insulators suffer from issues realted to industrial compatibility [\onlinecite{pai2018switching}], while heavy metals and alloys have limitations with respect to the spin diffusion length due to the spin relaxation being controlled by the Elliot-Yafet [\onlinecite{zimmermann2012anisotropy}] and Dyakonov-Perel [\onlinecite{long2016strong}] scattering mechanisms. The conducting TMD 1T-TaS$_2$ investigated in this work is an industrial compatible material as well as easy to fabricate for SOT devices and therefore avoid such limitations. Moreover, 1T-TaS$_2$ provides rich physics due to its inherent property of charge density wave fluctuations where electrons collectively may carry a charge current in a highly correlated fashion.

\begin{figure*}[h!tbp]
\centering
\includegraphics[width=17cm]{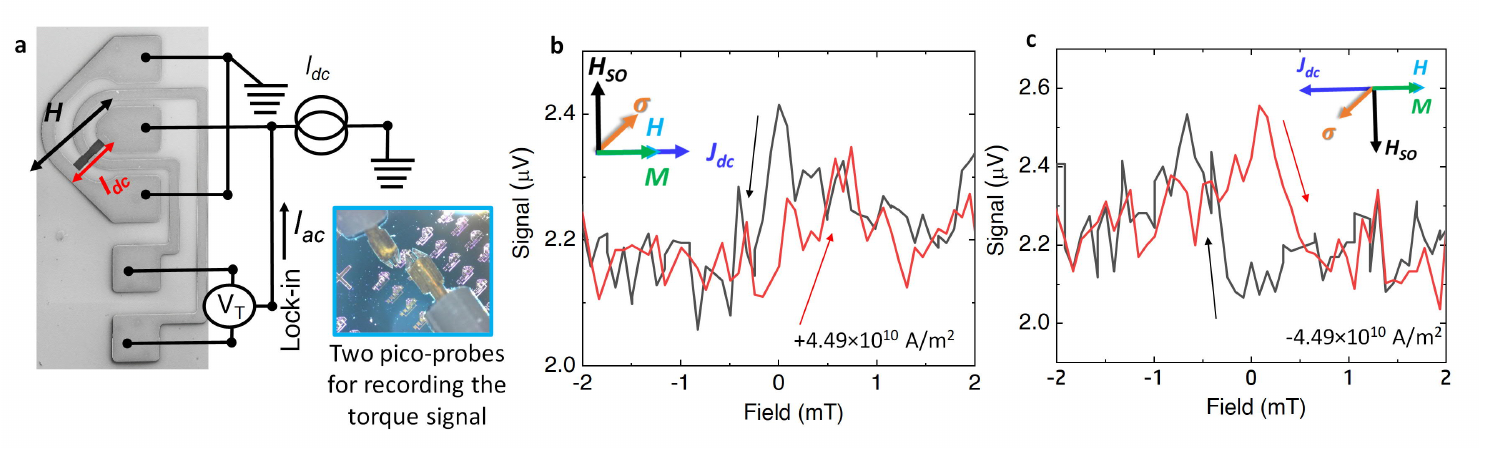}\\
\caption{\textbf{a,} SEM image of device together with schematic of the measurement circuit for transverse voltage measurement in the in-plane field geometry. Two pico-probes were used for injecting a charge current and recording the voltage signal simultaneously. Hysteresis loops recorded ifor positive and negative dc-current; \textbf{b,} +4mA , and \textbf{c,} -4mA. Schematics in panels b and c above the hysteretic plots show the directions of torques acting on the magnetization for the positive and negative current biasing.}\label{fig:fig4}
\end{figure*}

The dc-current induced changes of the effective damping can also be seen in the  $\mu_0 \Delta H$ versus $f$ results shown in Fig. 3(c) for positive and negative dc-currents. The current distribution in the heterostructure was evaluated and it was found that 19\% of the current is flowing through the 1T-TaS$_2$ layer (see Supplementary information S8). Consequently, the Oersted field $\mu_0 H_{Oe}$ in the 1T-TaS$_2$ layer is found to be $\sim$0.012mT/mA, which is very small, and it is therefore concluded that the field-like torque contribution generated by dc-current passing through TaS$_2$ layer can be neglected. The ST-FMR spectra and resonance field plots shown for two currents in Figs. 3(d) and 3(e), respectively, show no change during current polarity reversal, which is indicative of negligible field-like torque contributions. However, a small field-like torque contribution can arise from the unavoidable interface symmetry breaking [\onlinecite{amin2016spinforma},\onlinecite{amin2016spinpheno}], which is discussed in Supplementary information S10.

\section{Validation of dc-current induced damping-like torque in T\lowercase{a}S$_2$/P\lowercase{y} device}

\textit{\textbf{Magnetization switching in presence of in-plane field at constant dc-current}}
In presence of a charge current, the magnetization dynamics is described by the modified Landau-Lifshitz-Gilbert equation and follows the Slonczewski model where the additional torque is added due to the spin-orbit interaction (SOI) induced SHE and can be expressed as [\onlinecite{liu2011spin},\onlinecite{slonczewski2002currents}], $\frac{d{m}}{dt}=-\gamma {m} \times {H_{eff}}+\alpha {m} ({m} \times {H_{eff}})+\tau_{DL}({m}\times({\sigma} \times {m}))-\tau_{FL}({m} \times {H_t})$.
Here, the first and second terms account for the precessional motion and relaxation of the magnetization towards the equilibrium direction, respectively. The third and fourth terms represent the damping-like and field-like torques, respectively, induced by the charge current. These torques are orthogonal to the normalized magnetization $\boldsymbol{m}$, and $\tau_{DL}(=\gamma \frac{\hbar \theta_s^\alpha}{2e \mu_0 M_s t_{Py}}J_{C,dc})$ and $\tau_{FL}$ are the damping-like and field-like torque weight factors, respectively. The total field is given by $H_t=H_{rf}+H_{Oe}$, where $H_{rf}$ and $H_{Oe}$ are the Rashba-Edelstein and Oersted field contributions, respectively. Symbols $\sigma$, $\gamma$, $H_{eff}$ and $J_{C,dc}$ are used as notations for the spin polarization, gyromagnetic ratio, effective field and charge current density, respectively.
The damping-like torque acts as an out-of-plane magnetic field, $H_{SO}$ [\onlinecite{khang2018conductive}], whose amplitude can be evaluated from $\frac{\tau_{DL}}{\gamma}(=\frac{\hbar \theta_s^\alpha}{2e \mu_0 M_s t}J_{C,dc})$. Using the values for the charge-spin conversion efficiency (0.25), dc-current density (4.49 $\times$10$^{10}$A/m$^2$ at 4 mA dc-current) and the saturation magnetization (820 kA/m) of Py, the value of $\tau_{DL}/\gamma$ is found to be $\sim$324 A/m. Here, the transverse voltage signal has been measured in an in-plane magnetic field at two (positive and negative) dc-currents (see Supplementary information S1 for details of the measurement). Figures 4(b) and 4(c) show the voltage signal recorded at positive and negative current densities, respectively. At +4.49$\times$10$^{10}$ A/m$^2$, a traditional hysteresis loop is formed while changing the dc-current to -4.49$\times$10$^{10}$A/m$^2$, both $\sigma$ and H$_{SO}$ reverse directions yielding an inverted hysteresis loop, which is in consonance with the behaviour of the SOT [\onlinecite{khang2018conductive},\onlinecite{li2018spin},\onlinecite{fan2013observation}]. This magnetization switching indicates a sizable damping-like toque in Py/TaS$_2$. The Hall voltage hysteresis was also recorded using a Hall bar structure as is shown in Supplementary information S11. Further, using micromagnetic simulations, the switching of the in-plane magnetization for different current amplitudes has also been studied (Supplementary information S12), which supports the observation of inverted hysteresis loops in experiments.

\textit{\textbf{Angular dependent Planar Hall Effect (PHE) for spin-orbit torques}}
Planar Hall effect (PHE) measurements have received much attention for characterizing the spin orbit torques (SOTs) in the in-plane magnetized systems [\onlinecite{fan2013observation},\onlinecite{mahendra2018room},\onlinecite{kawaguchi2013current}]. Figure 5(a) shows a scanning electron microscope (SEM) image of the Hall device used for planar Hall effect measurements. A schematic of the measurement circuit has been added to the image. In the planar Hall measurement, the sample is rotated 360 degrees at fixed dc-current. The vector representation of the planar Hall effect is shown in Fig. 5(b), where $\varphi$ is the angle between the current direction and the applied field and $\theta$ is the angle between the current direction and the magnetization vector. The theoretical background of the planar Hall effect is discussed in Supplemenetary information S11. Figure 5(c) shows the planar Hall effect signal ($R_H$) versus $\varphi$ recorded for two dc-currents of same magnitude but of opposite polarity. A magnetic field of 0.4 T was used for the measurements, which was enough for suppressing the field-like torque contribution. 
\begin{figure}[b]
\includegraphics[width=8cm]{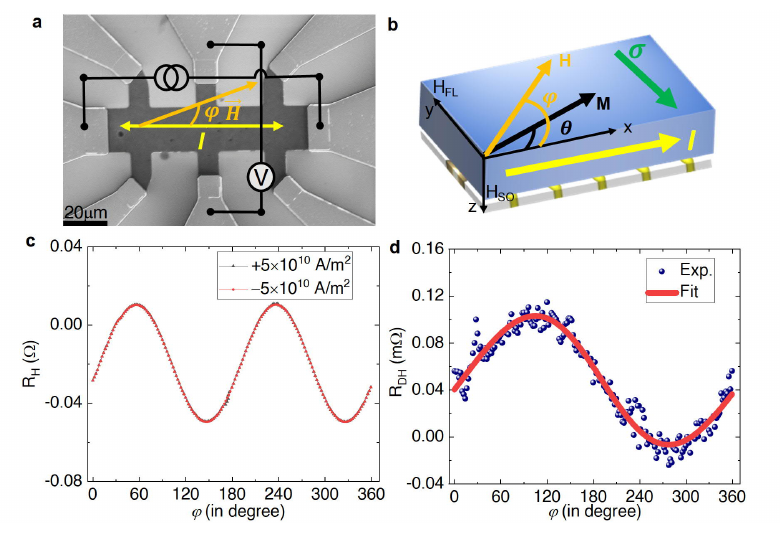}\\
\caption{\textbf{a,} SEM image of Py/TaS$_2$ Hall bar device together with a schematic of the circuit used for planar Hall effect measurement. The angle $\varphi(\theta)$ is the in-plane angle between the magnetic field (magnetization) and the current direction in the device. \textbf{b,} Vector components of the applied field H , magnetization and the fields generated by SOTs. \textbf{c,} Planar Hall effect signal recorded for two directions of the dc-current (magnitude=5$\times$10$^{10}$A/m$^2$) in presence of an in-plane magnetic field of 0.4 T. \textbf{d,} Plot of the Hall resistance difference observed for two current directions versus angle $\varphi$.}\label{fig:fig5}
\end{figure}
The planar Hall Effect measured at different magnetic fields and currents are discussed in the Supplementary information S11. The difference between the curves ($R_{DH}$) is plotted as a function of $\varphi$ in Fig. 5(d), which embraces the dominance of the damping-like torque. The $R_{DH}$ versus $\phi$ curve was fitted using Eq. S12 (see Supplementary information S11) with $H_{FL}$ and $H_{SO}$ as fitting parameters. The $H_{SO}$ field determines the amount of damping-like torque acting on the ferromagnetic layer. The value is found to be 2.5 Oe per 10$^{10}$ A/m$^2$, from which the spin torque efficiency ($\zeta_S$) has been obtained by using the expression, $\zeta_S=(2e M_s t_Py)/\hbar  H_{SO}/J_c$ . The field-like contribution $H_{FL}$ has been found to be -0.02 Oe per 10$^{10}$ A/m$^2$. The dc-current density $J_c$ can be calculated by considering the dimensions of the device and the conductivity values at a dc-current value of 4.5 mA,  yielding a dc-current density of 5 $\times$10$^{10}$ A/m$^2$. Using the thickness and saturation magnetization of the Py layer, $\zeta_S$ is found to be 0.43 which is larger then the value obtained using the ST-FMR analysis. Therefore, it is confirmed without doubt that a monolayer of 1T-TaS$_2$ produces a damping-like torque acting on the Py magnetization.
\begin{figure*}[h!tbp]
\includegraphics[width=15cm]{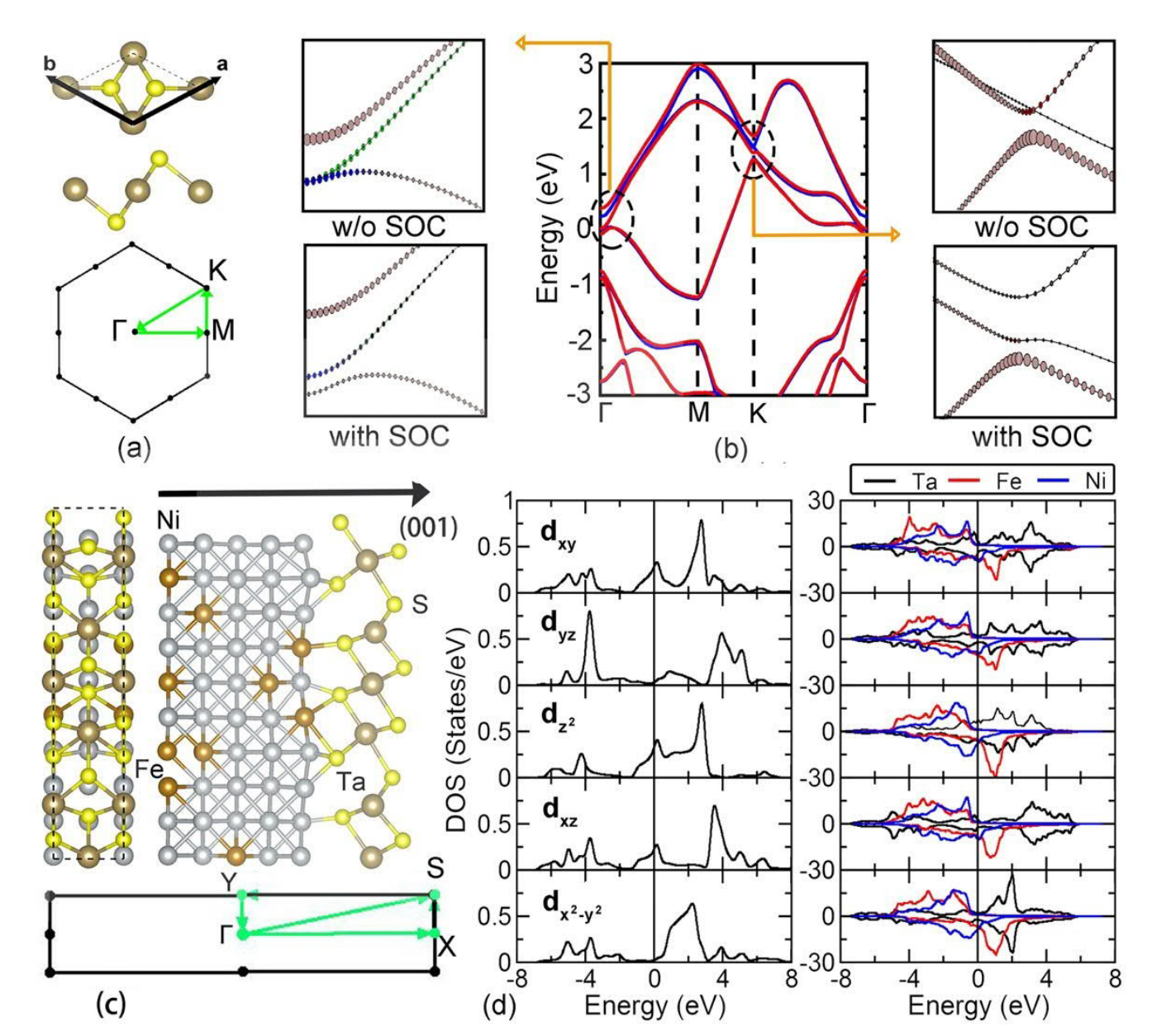}
\caption{\textbf{a,} Unit cell and Brillouin zone of pristine TaS$_2$. Both top and side views of the hexagonal unit cell are shown. \textbf{b,} Band structure of pristine TaS$_2$ with and without spin-orbit coupling (SOC) (middle panel), expanded view of a part of the band structure (encircled) at the $\Gamma$ point without and with SOC (left panel) and the same but at the K point (right panel). \textbf{c,} Top and side views of the unit cell of TaS$_2$/Py illustrating the optimized geometry. In the lower panel, the BZ is shown.  \textbf{d,} (left) Projected densities of states (DOSs) for different d-orbital symmetries of Ta for pristine TaS$_2$ and (right) spin-polarized projected DOSs for TaS$_2$/Py system where the d-orbitals of Ta, Fe and Ni are shown.}\label{fig:fig6}
\end{figure*}

\section{Discussion} 
It is to be pointed that the possibility of a finite field-like torque contribution is not ruled out (see Supplementary information S10) in this work, which is reasonable and cannot be disentangled in SOT based systems [\onlinecite{amin2016spinforma},\onlinecite{amin2016spinpheno},\onlinecite{kim2017spin}]. There has been no clear evaluation of the critical current density and spin torque efficiency in previously reported results for TMD/FM heterostructures [\onlinecite{guimaraes2018spin},\onlinecite{shao2016strong}]. The quantitative estimation of the spin Hall conductivity and auto-oscillations current density in our Py/TaS$_2$ hold valuable information for several spintronic applications. Evidently, the interface of our Py/TaS$_2$ bilayer, as confirmed by cross-sectional TEM  and supported by parameters extracted from X-ray reflectivity measurements, is clean in contrast to other works using exfoliated sheets and non-uniform growth where extrinsic contributions from strain and defect related issues [\onlinecite{guimaraes2018spin},\onlinecite{xu2020high},\onlinecite{zibouche2014transition}] reduce the charge-spin conversion efficiency. 

First principles calculations based on density functional theory reveal the role of SOC for lifting the degeneracy in the band structure of TaS$_2$/Py. Figure 6 shows the energy band structures without and with the inclusion of SOC for both pristine TaS$_2$ and Py/ TaS$_2$ systems. A detailed discussion of the structural and electronic properties is presented in the Supplementary information S2. For pristine TaS$_2$, the effects of SOC are clearly observed at $\Gamma$ and K points (see the expanded views). Specifically, at the $\Gamma$ point, the degenerate d$_{xz}$ and d${_{yz}}$ bands are split due to SOC. It should be noted that this degeneracy is already lifted by the lower symmetry present at the the interface between TaS$_2$ and Py due to the distorted atomic structures. On top of that, further splitting occurs due to the presence of SOC. Therefore, one can conclude that a sizeable redistribution of band structure and hence the splitting of states due to the interface occurs, which becomes responsible for a prominent damping like torque.  To highlight the contribution from different d-orbitals, we show in Fig. 6(d) the orbital projected DOSs of Ta in pristine TaS$_2$ and also for the TaS$_2$/Py bilayer. Moreover, projected DOSs of d-orbitals of Fe and Ni in Py are shown to reveal features of hybridization. As our energy range of interest is in the vicinity of the Fermi level, we will consider the electronic states within that energy range. It is observed that in the pristine material, d$_{xy}$, d$_{z^2}$ and d$_{xz}$ are the orbitals of interest. However, for the bilayer, the d$_{z^2}$ orbital for both spin channels is quite prominent at the Fermi level and its vicinity. Moreover, for the spin-down channel, a hybridization between the d$_{z^2}$ orbitals of Ta and Ni is seen for the spin-down channel.

\section{Conclusions}

In conclusion, the damping-like spin torque efficiency has been carefully investigated in the Py/TaS$_2$ bilayer using spin-torque ferromagnetic resonance and Hall effect measurements. Employing effective damping modulation or changes with dc-current, the effective spin torque efficiency is found to be 0.25$\pm$0.03. 
Angle dependent planar Hall effect measurements verify the spin-torque efficiency and clearly reveal the dominance of damping-like torque in our TaS$_2$/Py bilayer. Further, the microscopic origin of the observed dominance of damping-like torque has been substantiated by DFT calculations. The observation of a dominating damping-like torque in just one monolayer provides a path for how to use TaS$_2$ in future spin-orbitronic devices.

\begin{acknowledgments}
The Swedish Research Council (\,VR)\, supports this work, grant no: 2017-03799. Authors thank Seda Ullusoy for SEM imaging. We thank CWT for providing cross section TEM measurements.
\end{acknowledgments}

\bibliography{apssamp}
\bibliographystyle{apsrev4-2}
\end{document}